# Small-scale polarization variations in near-inertial deep Mediterranean horizontal waterflow

**by Hans van Haren**

Royal Netherlands Institute for Sea Research (NIOZ), P.O. Box 59, 1790 AB Den Burg, the Netherlands.
e-mail: hans.van.haren@nioz.nl

Deep-sea observations are reported of horizontal waterflow differences with typical amplitudes of 0.02 m $s^{-1}$ over 50-m small scales so that relative vorticity reaches values of the inertial frequency f. The timeseries observations are made using a complex mooring system deployed in the 2500-m deep Mediterranean Sea, where vertical density stratification is extremely weak, with buoyancy frequency O(f), and the slow waterflow with total speeds <0.07 m $s^{-1}$ is dominated by inertial internal waves and sub-mesoscale eddies. Horizontal waterflow differences increase when polarization, i.e. direction of traversal of elliptic oscillatory motion, switches sign. Common anticyclonic polarization of inertial motions is predominantly found under near-homogeneous conditions. It alternates with uncommon cyclonic polarization under stratified-water conditions, varyingly over 50-m distances. The alternation is in line with predictions from non-traditional inertio-gravity wave theory, but only when relatively strong turbulent convection causes local reduced stratification, as observed.

## 1. Introduction

While in hydrodynamic laboratory models flow conditions may be controlled in a limited number of directions and scales, such is not possible in geophysical flows. The study of geophysical flows involves motions at many scales distributed over about 10 orders of magnitude and requires elaborate data processing to isolate specific conditions. This is because geophysical flows may be dominated by some specific motions, such as tides, but in general demonstrate nonlinear interactions of various degrees between different motions like eddies and waves that are difficult to separate. Most complex are turbulent motions, with scales varying from 0.001 to 100 m large overturns in the deep sea.

A standard model for energy transfer between turbulent motions is unidirectional from large to small scales within the 'inertial subrange' (Kolmogorov 1941). However, ambiguous directions are found for energy transfer between eddies and turbulence, in laboratory models (Lohse & Xia 2010) as well as in geophysical data (Bolgiano 1959; van Haren 2025), but to unknown extent. Nonetheless, these partially unknown interactions between different motions are vital for life, e.g. in the atmosphere and ocean.

One of the unknowns in open-ocean flows is the interaction of mesoscale O(10-100) km two-dimensional eddy flows with sub-mesoscale O(0.1-10) km eddies and internal waves that may lead to three-dimensional turbulence and eventually irreversible mixing of suspended matter. Vertical waterflow differences and scales have been regularly studied in the stably stratified ocean environment on the importance of destabilizing shear, e.g. using multiple instrumented single mooring lines. However, horizontal scales are not often studied and multiple mooring lines are seldom deployed at scales of <10 km in the deep ocean, rare exceptions being Briscoe (1975) resolving open-ocean scales down to O(1-1000) m and Saunders (1983) resolving deep-ocean scales down to O(10) km.

As earth rotation plays an important role for geophysical flows that have timescales exceeding $1/f$, $f$ the local inertial frequency, one expects that it also affects the waterflow differences and nonlinear interactions. Oscillatory flows such as at tidal and inertial frequencies are not channel-like rectilinear, but more commonly elliptic in the unbounded open ocean. Any oscillatory horizontal motion on a rotating sphere can be decomposed into four components (Gonella 1972; Prandle 1982): 1) major axis amplitude, 2) eccentricity or minor over major axis amplitude ratio, 3) phase of ellipse traversal with respect to a given moment in time, and 4) inclination or orientation of the ellipse. To these, one may add a fifth component: 5) polarization, the direction in which the ellipse is traversed, which is contained in the sign of eccentricity. A switch in polarization implies a reversal of ellipse traversal.

Differences in waterflow, in the vertical or horizontal, may be brought about by variations in any of these five components. The classic vertical shear is via oppositely directed steady flows of equal magnitude. However, oscillatory flows may also set-up vertical shear via a change in amplitude, phase and/or inclination (Maas & van Haren 1987; van Haren 2000). In the ocean, shear across stratification is dominated by near-inertial motions because of their small vertical scales, possibly leading the scale-limit O(10) m of vertical shear (Gargett *et al*. 1981), and because shear is always maximal, i.e. oppositely directed flow, also for near-circular oscillatory motions at any given time.

Frictional oscillatory flows like barotropic tides in shallow seas may demonstrate vertical polarization switches in the interior (Maas & van Haren 1987), but the effects on vertical shear were not established. In contrast, polarization switches are not expected for freely propagating internal waves in uniformly stratified waters, which classically under the traditional approximation show anticyclonic polarized, clockwise motion in the Northern Hemisphere (e.g., Gonella 1972).

Nevertheless, rare polarization switches with time have been observed for internal tides under varying turbulence and small-scale shear conditions in a weakly stratified deep North-Atlantic channel (van Haren 2015). These polarization switches were hypothesized to reflect non-traditional inertio-gravity waves IGW (Gerkema & Exharchou 2008), which are an extension of the classic internal wave band (LeBlond & Mysak 1978). The importance of short-scale variations in polarization for interaction between vertical shear and stratification, via turbulent mixing, raises interest in observations of horizontal variations in ellipsoidal oscillatory motions that are quite unknown.

In this paper, to test the hypothesis of polarization switches inducing horizontal waterflow differences, they are investigated over short 50-m scales measured at about 125 m above a flat seafloor in 2500-m deep Northwestern Mediterranean waters (Fig. 1a). At this site, internal wave motions are dominated at f, not at tidal frequencies, and stratification is very weak so that the average buoyancy frequency N = O(f) (van Haren *et al*. 2026). The technical challenge was to deploy mooring lines at such close distance at great depths and to resolve waterflow-difference amplitudes of <0.01 m $s^{-1}$ where maximum flow speeds are <0.07 m $s^{-1}$. The focus is on the observation of sign of polarization and of hypothesized potential effects on flow interactions. Theoretically for example, varying the tilting of an IGW beam, the cone of propagation (LeBlond and Mysak 1978), under varying stratification may lead to ellipticity variations. In the horizontal such variations between motions may lead to flow divergence and relative vorticity, which affects nonlinear flow interaction and redistribution of suspended matter.

**2. Theory**

One of the potential effects of buoyancy-driven turbulent convection is the development of two-dimensional '2D' eddies, which also associate with frontal collapse, and which have horizontal scales O(0.1-10) km in the ocean: sub-mesoscales (e.g., Taylor & Thompson 2023). (While eddies are characterized here as 2D rather than 3D because of their small aspect ratio << 1, it is acknowledged that they can have stronger flows in more stratified waters near the surface than in the deep sea.) Sub-mesoscale eddies are found ubiquitous in the Northwestern Mediterranean Sea (Gascard 1978; Testor & Gascard 2006) and are suggested to interact with large-scale features such as continental boundary flows. In this region, other mechanisms may lead to sub-mesoscale eddies as well. Such occurs for example, when rare deep dense water is formed, and, much more commonly, when the boundary flows spin off eddies at mesoscales with typical 10-20 day periodicity (Crepon *et al*. 1982; van Haren *et al*. 2011), which may break up to sub-mesoscales.

The region also shows extensive generation of near-inertial waves by winds from large mountain ranges like the Alps and the Pyrenees, which, under conditions of homogeneous and weakly stratified waters, may lead to slantwise-to-vertical convection, as has been proposed in general for convection plumes mainly from models (Marshall & Schott 1999; Straneo *et al*. 2002; Sheremet 2004) and inferred from shipborne profiling observations in the open Western Mediterranean (van Haren & Millot 2009). Slantwise-homogeneous convection may develop to irreversible 3D turbulence, while vertical density profiles appear stably stratified (Straneo *et al*. 2002).

Slantwise convection is related with the extension of the internal wave band into the sub-inertial, sub-mesoscale frequency range. Under weakly stratified conditions, the traditional internal-wave bounds [f, N] extend to IGW bounds [$\omega_{min}$ $\omega_{max}$] (LeBlond & Mysak 1978). Here, planetary local vorticity $f = 2\Omega\sin\varphi$, in which $\varphi$ denotes latitude and $\Omega$ the Earth rotational frequency. The $\omega_{min}<f$ and $\omega_{max}>2\Omega$, when $2\Omega > N$, are functions of N, $\varphi$ and direction of wave propagation (LeBlond & Mysak 1978; Gerkema *et al*. 2008),

$$\omega_{max}, \omega_{min} = (D \pm(D^2-E^2)^{1/2})^{1/2}/\sqrt{2}, \qquad (2.1)$$

in which $D = N^2 + f^2 + f_s^2$, $E = 2fN$, and $f_s = f_h\sin\alpha$, $f_h = 2\Omega\cos\varphi$ and wave-direction angle $\alpha$ to the zonal. For $f_s = 0$, i.e. latitudinal propagation, or $N >> 2\Omega$, i.e. strong stratification, the classic [f, N] follow from (2.1).

Internal-wave bounds may also vary due to (sub-)mesoscale motions. Local time- and space-varying differences in horizontal waterflow (u, v), such as occur in meanders and eddies, can generate relative vorticity $\zeta = \partial v/\partial x - \partial u/\partial y$, so that the effective inertial frequency reads (Kunze 1985),

$$f_{eff} = (f^2 + f\zeta - \partial u/\partial x\cdot\partial v/\partial y)^{1/2}. \qquad (2.2)$$

For weak $\zeta < 0.2f$, (2.2) can be approximated by $f_{eff} \approx f + \zeta/2$ (Mooers 1975; Kunze 1985), but this is probably not relevant for the Northwestern Mediterranean where $|\zeta| = f/2$ are reported from mid-depth drifter observations (Testor & Gascard 2006). When $f_{eff} < f$, relative vorticity is dominated by anticyclonic motions, while for $f_{eff} > f$ cyclonic motions dominate.

### *2.1 IGW polarization switches*

One of the characteristics contrasting IGW from traditionally modelled internal waves is that they have multiple directionalities instead of a single one (e.g., LeBlond & Mysak 1978; Gerkema *et al*. 2008). Theoretically, the upgoing $\xi_+$ and downgoing $\xi_-$ beams of IGW-energy may bring about polarization switches (Gerkema & Exarchou 2008), like observed for tropical deep-ocean internal tidal currents (van Haren 2015). For IGW-motions, $\xi_\pm = \mu_\pm x - z$, and steepness reads (e.g., Gerkema et al., 2008),

$$\mu_\pm = (B \pm (B^2 - AC)^{1/2})/A, \quad (2.3)$$

in which $A = N^2 - \omega^2 + f_s^2$, $B = ff_s$, and $C = f^2 - \omega^2$. As demonstrated by Gerkema and Exarchou (2008), the polarization is determined by the sign of,

$$f_s\mu_\pm - f, \quad (2.4)$$

and is cyclonic when (2.4) is positive. In the Northern Hemisphere only $\mu_+$ satisfies the polarization switch condition $f = f_s\mu_\pm$ describing a rectilinear motion. The more horizontal propagation $\mu_-$ always shows anticyclonic polarization for near-inertial motions, like under traditional approximation and for latitudinal propagation. (For the Southern Hemisphere the characteristics of up- and downgoing beams should be interchanged.)

Following theoretical modelling by Gerkema and Exarchou (2008), under a number of assumptions including unbounded ocean with time-independent constant N and $f_{eff} = f$ so that relative vorticity is negligible, $\mu_+$ at given frequency $\omega$ may switch polarization as a function of N and $\varphi$ at the rectilinear-motion condition,

$$\omega = fN/(f^2+f_s^2)^{1/2}, \quad (2.5)$$

in the Northern Hemisphere. For example, for just super-inertial $\omega >\sim f$ and meridional propagation $\alpha = \pi/2$, a switch between cyclonic and anticyclonic polarization is possible when,

$$N \approx 2\Omega. \quad (2.6)$$

For $\alpha << \pi/2$, (2.6) approaches the inertial frequency, $N >\sim f$. In practice also in the deep sea however, stratification and relative vorticity are not constant, but space and time dependent. As a result, one can expect variations in polarization sign from moored instruments in varying, weakly stratified waters.

## 3. Materials and Methods

### *3.1 Site and instrumentation*

A multiple vertical-lines 'large-ring mooring' was deployed at the <1° flat and 2458-m deep seafloor of 42° 49.50′N, 006° 11.78′E just 5 km south of the foot of the steep continental slope of the Northwestern Mediterranean Sea (Fig. 1). Details of mooring layout and deployment are given in van Haren *et al.* (2021). In summary, 45 vertical lines 125-m tall were each tensioned to 1.3 kN by a single buoy on top. The lines mainly held temperature sensors, of which only vertical temperature differences are discussed in this paper. Three buoys held a 2-MHz single-point Nortek AquaDopp acoustic current meter recording samples of waterflow and relative acoustic backscatter amplitude at a rate of once per 600 s. As a diagnostic for turbulence, acoustic amplitude variations are used, which correlate significantly without phase-lag with vertical temperature variations that correspond positively with turbulence intensity (van Haren *et al.* 2026). Largest turbulence due to slantwise convection from above is found when waters are relatively stratified in the vertical, so that $N \approx 2f > 2\Omega$. In thin layers of $\Delta z = 10$ m, $N = 4f$ may occur. Such periods are alternated with lesser turbulent waters that are near-homogeneous $N \approx 0.5f$ in the vertical (van Haren *et al.* 2026). As a result, polarization switches are expected from (2.5).

The current meters were at the same height above the seafloor, h = 126.0±0.4 m. Their mutual horizontal separation was $\Delta_h = 50 \pm 1.5$ m in an equilateral triangle, with the position of line 14 closely pointing northward (Fig. 1b). The nominal instrumental precision of horizontal waterflow is 0.009 m $s^{-1}$ for each sample, so that noise-filtering with cut-off at 4 cpd (cycles per day) reduces the precision or one standard deviation error to 0.0015 m $s^{-1}$. Relative acoustic backscatter amplitude dI = (amp - $I_{ref}$)*0.45 [dB]. Raw amplitudes 'amp' are relative to reference level $I_{ref}$. To represent volume backscattering, dI are transferred to relative echo intensities rEI = $10^{dI/10}$ with arbirary power units.

### *3.2 Data processing for polarization determination*

After writing horizontal oscillatory flow components u(t) v(t) in a plane with real and imaginary axes as $u - iv \approx A\exp(-i\omega t+\theta)$, $i^2 = -1$, A amplitude, ω frequency, and θ phase, the time-dependent polarization P is inferred as (Gonella 1972),

$$P(t) = -i\,\mathrm{dlg}(u - iv)/dt, \quad (3.1)$$

where lg mathematically denotes the logarithm with base 10. P is positive for cyclonic motions and negative for anticyclonic motions. Here, we are only interested in the sign *sgn*(P) of the polarization and not in the amplitude-phase modulation.

As the dominant oscillatory motions at the deep Mediterranean site are not barotropic tidal but baroclinic near-inertial waves, the frequency of motions is not a single sharp deterministic one but distributed over a finite band $\omega_f \approx (1.03\pm0.1)$f. As will be demonstrated in Section 4, internal-wave energy can be transferred to more IGW frequencies. Therefore, (3.1) is not applied to a single-frequency motion, for which *sgn*(P) would be fixed anyways, but to motions in finite-width bands.

Three bands are defined around f and involving IGW using rather sharp double-elliptic phase-preserving filters (Parks & Burrus 1987): i) near-inertial band 0.87f < ω < 1.13f, ii) IGW-$N_{max}$ (maximum 2-m-small-scale buoyancy frequency) band 0.8 < ω < 4 cpd (the noise cut-off), which includes IGW, for mean N = f, and motions at frequencies f, 2Ω, and 2f, and iii) sub-mesoscale-IGW band 0.4 < ω < 2 cpd, which includes IGW, for N = 0.5f under near-homogeneous conditions, f, 2Ω, and sub-mesoscale motions. The latter band only partially resolves sub-mesoscales, which are defined between 0.1 < ω < $\omega_{min}$. Mesoscales are defined for motions at ~0.03 < ω < 0.2 cpd.

## 4. Results

### *4.1 Overview*

A two-month late-autumn timeseries demonstrates considerable variation in waterflow (Fig. 2a) and vertical temperature difference and relative echo intensity (Fig 2b) amplitudes. Near-homogeneous conditions are defined when 124-m vertical temperature difference $\Delta\Theta < \Theta_{thr} = 2\times10^{-4}$ °C. Vertical stratified-water conditions apply when $\Delta\Theta > \Theta_{thr}$. Besides variations at near-inertial IGW timescales, variations are observed at (sub-)mesoscales of 2-15 days. This results in variations in local effective vorticity, governed by relative vorticity O(f) in both cyclonic and anticyclonic directions, as computed at 50-m horizontal distances (Fig. 2c). Waterflow amplitudes seldom exceed 0.06 m $s^{-1}$, which indicates that they not associate with vigorous convection processes like deep dense-water formation requiring speeds >0.2 m $s^{-1}$ and which occur mainly in winter, not in autumn (e.g., van Haren *et al.* 2011).

In Fig. 2, near-homogeneous conditions occur for example around days 320 and 353, while stratified -water conditions around days 330 and 343. Coarsely, near-homogeneous conditions associate with mean flow speed $|U| \approx 0.02$ m $s^{-1}$ and small relative vorticity, i.e. $|f_{eff} - f|/f < 0.5$, whereas stratified-water conditions associate with $|U| \approx 0.04$ m $s^{-1}$ and large relative vorticity, i.e. $|f_{eff} - f|/f \approx 1$ (Fig. 2c). Such large values for relative, horizontal waterflow-differences induced vorticity confirms mid-depth observations by Testor and Gascard (2006) in the area. However, the sign of the sub-mesoscale relative vorticity does not seem to determine the levels of turbulence and stratification.

Averaged over the two months of data, the kinetic energy spectrum (Fig. 2d) demonstrates the horizontal white noise level for $\omega > 4$ cpd, a gradual slope of power-law for $\omega < 4$ cpd, interrupted by a peak at $\omega = (1.03\pm0.12)f$ of more than one order of magnitude larger variance than at surrounding frequencies. Fig. 2d-f also display the three filter passbands, with the sub-mesoscale-IGW band being cut-off close to a change in slope around 0.4 cpd $\approx \omega_{min}(N=0.5f)$. Two 11-day averaged spectra represent near-homogeneous conditions dominated by inertial motions (Fig. 2e) and stratified-water conditions that show increased kinetic energy at all frequencies except at f (Fig. 2f). Under stratified-water conditions, the increased ‘noise’ levels pass into IGW, which suggests an indistinguishable transition from turbulence into IGW, albeit poorly resolved. The integrated enhanced IGW motions reach similar speeds of 0.02 m $s^{-1}$ as those of the near-inertial band of Fig. 2e, which may suggest a transfer of near-inertial motions to a much broader IGW under stratified-water conditions.

### *4.2 IGW model prediction of polarization switches*

Adapted for observed near-inertial frequencies under near-homogeneous conditions and the stratification range up to stratified-water conditions, the IGW-model (Gerkema and Exarchou, 2008) for $\mu_+$ and arbitrary $\alpha = \pi/3$ is given in Fig. 3 for $\max(\omega_{min}) < \omega < \min(\omega_{max})$. While the midrange $N \approx f$ provides anomalous cyclonic polarization at all IGW frequencies, two possible transitions to anticyclonic polarization are predicted. A smooth transition occurs for $N > 1.2f$, linearly increasing with $\omega$, being nearly 1.4f around $\omega = f$ and thus representing (2.6) for 60°-from-latitude propagation. A sudden transition across a singularity occurs for $N < 1.0f$ and $\omega > 0.93f$. The parameter range of near-inertial motions observed in the deep Mediterranean under near-homogeneous conditions lies across this sudden transition, while that observed under vertical stratified-water conditions is predicted to be always anticyclonic polarized. Recall that horizontal IGW motions for $\mu_-$ are always anticyclonic polarized.

### *4.3 Observations of polarization switches*

Over an 11-day period the transition from near-homogeneous to stratified-water conditions around day 325 (Fig 2b) is evidenced in single current meter data as a distinct switch in near-inertial polarization sign (Fig. 4). While the overall noise-filtered data show *sgn(*P) switches that have some association with sub-mesoscale switches (Fig. 4a), the near-inertial band shows a single switch from anticyclonic polarization to cyclonic polarization at day 326.0 (Fig. 4b). Broadening the band-pass filter to sub-mesoscale-IGW motions leaves the same impression of polarization switch at day 326.0 but adds a short switch at day 323 and multiple ones after day >329 (Fig. 4c).Widening the IGW band to include small-scale stratified motions shows the most variable switching of polarization, albeit still mostly being anticyclonic for days <326 and cyclonic for days >326 (Fig. 4d).

Following IGW-model, common near-inertial IGW motions are to have anticyclonic polarization in the near-horizontal propagating downgoing beam, but ambiguous, albeit mainly cyclonic, polarization under near-homogeneous conditions and anticyclonic polarization under stratified-water conditions in the upgoing beam (Fig. 3). This is partially explaining the observed anticyclonic motions under near-homogeneous conditions also for the upgoing beam, provided $N < 0.4f$. However, it can only explain the cyclonic polarization under vertical stratified-water conditions $N = 2f$ if the actual convection is slantwise directed, so that stratification in the direction of convection measures the equivalent of vertical $0.4f < N < 1.2f$, for $\omega \approx f$ (Fig. 3). If $f_{eff} \approx 0$, cyclonic IGW-motions are predicted for $0 < N < 0.7f$, for $\omega < f$ (not shown).

The potential impact of these observed polarization switches may be inferred from Fig. 5. The noise-filtered waterflow amplitudes from all three current meters show general agreement well in phase of dominant inertial periodicity e.g. around day 321, sudden deviations of one instrument e.g. around day 324, and multiple mutual differences and erratic periodicities for days >326 (Fig. 5a). This change in waterflow amplitudes associates with the general switch in polarization sign at day 326 (Fig. 5b). As a consequence, the amplitudes of horizontal waterflow differences vary from near-zero for days < 322, via about 0.02 m $s^{-1}$ change over an inertial period for days 322-326, to irregularly attaining values >0.02 m $s^{-1}$ for days >326. The 50-m waterflow differences reach up to the value of the local mean speeds, and reflect substantial variations in eddy and convective turbulent motions. These variations all significantly exceed instrumental noise levels of about 0.002 m $s^{-1}$ (Fig. 5c-f).

The u-v-hodographs dominated by near-inertial motions in Fig. 5d during an episode of near-homogeneous conditions show anticyclonic polarization at all three instruments, and nearby start- and end-points indicating similar amplitudes and phase. The ellipse ratio of short over long axes is 0.4-0.5, which approximates the theoretical value for quasi-gyroscopic waves at the local latitude (van Haren and Millot 2004; Gerkema et al. 2008). Small variations are observed in ellipse inclination. In addition, near-inertial ellipses are modified, presumed by weakly nonlinear motions that incur short switches in polarization at two current meters, resulting in horizontal flow-difference amplitudes of <0.01 m $s^{-1}$.

From day 326 onward near-inertial horizontal waterflow amplitudes decrease and polarization switches are more common. In Fig. 5e, ellipses are still discernible at all three current meters, but one shows dominant cyclonic polarization. This leads to waterflow difference amplitudes of up to 0.02 m $s^{-1}$, despite the decrease in general near-inertial amplitudes. Further in time, in Fig. 5f the u-v-hodographs attain irregular butterfly-like patterns, all instruments reveal short and long polarization switches, and amplitudes of horizontal waterflow difference over 50-m distance peak at >0.03 m $s^{-1}$, which is close to the mean local speed of dominant eastward waterflow during the displayed period and well exceeds the amplitude of oscillatory motions.

However, the difference in hodograph's endpoints after one inertial period remains the same, about two standard-error deviations, as during earlier episodes at days <326. This is somewhat surprising because, in contrast with dominant inertial motions under near-homogenous conditions, vertical stratified-water episodes like during days 326-336 are dominated by enhanced turbulent convection that is slanted from above and sideways, as demonstrated in movies (van Haren *et al*. 2026). Given that observation, presumed near-homogeneity in slanted direction of convection tubes may explain the here observed frequent anticyclonic/cyclonic polarization switches for the entire IGW band, which has much more distributed kinetic energy to both $\omega_{min}$ and $\omega_{max}$ away from f (Fig. 2f).

## 5. Discussion

The deep Mediterranean Sea is characterized by distinct variations of vertical density stratification conditions between near-homogeneous and stratified-water, which alternate with mesoscale periodicity. To the characters of stratification conditions, being $N < 0.5f$ and $N = 2f$, kinetic energy peaking at f and distributed over IGW, relative sub-mesoscale vorticity so that $|f_{eff}-f|/f < 0.5$ and $|f_{eff}-f|/f \approx 1$, acoustic backscatter intensity near background levels and typically one order of magnitude larger, turbulence dominated by geothermal heating from below and three-times more energetic slantwise convection from above (van Haren *et al*. 2026), we can add the sign of horizontal near-inertial IGW waterflow ellipse polarization being predominantly anticyclonic and variable cyclonic/anticyclonic, respectively.

Although waterflow observations have been made at only one level of 125 m above seafloor, it is tempting to translate the observed horizontal flow differences to vertical variations. With typical 0.02 m $s^{-1}$ difference over 50 m one arrives at a value of $4\times10^{-4}$ $s^{-1}$ $\approx$ 4f under vertical stratified-water conditions. This is close to observed small-10-m scale variations in N (van Haren *et al*. 2026). Such small-scale, thin-layer stratification can support vertical current shear of similar value to attain marginal stability, also in the deep sea. This is then brought about by polarization switches in IGW, which, however, require nonlinear interaction for effective transfer to turbulence, e.g. via parametric instability or, more importantly, slantwise convection.

While the observations of polarization switches are distinct and clear, comparison with theoretical models is not straightforward under all conditions. From the IGW model for a downgoing beam, anticyclonic polarization is predicted under all conditions. Only the IGW model for an upgoing beam demonstrates sign-varying polarization for weakly stratified and near-homogeneous waters. Although the observed dominant anticyclonic polarization under near-homogeneous conditions may thus be explained for a downgoing beam, a transition to an upgoing beam is required under vertical stratified-water conditions. The IGW model for an upgoing beam also allows for anticyclonic near-homogeneous motions, under somewhat restrictive (ω, N) that, however, are very close to those observed. In contrast under vertical stratified-water conditions, polarization switching may occur between cyclonic and anticyclonic motions, but only around $N \approx 2\Omega$, which is $< 2f$ observed. To match with the model, a reduction in N is possible for relatively strongly varying conditions, like under vigorous turbulent convection as observed (van Haren *et al*. 2026), of which the plumes may turn slantwise, in which direction the stratification adopts the cyclonic/anticyclonic transition under near-homogeneous conditions $N \approx 0.4f$.

Although relative vorticity O(f) is suggested to explain the observed strong time-dependency of the frictional barotropic tidal-ellipse boundary layer (Maas & van Haren 1987), the contribution of similar sub-mesoscale vorticity to $f_{eff}$ is not expected to contribute much to polarization switches in the deep Mediterranean. For anticyclonic relative vorticity, so that $f_{eff} < f$, the IGW-model parameter space of polarization switches becomes more restrictive, with possible switches for $N < f$ and $\omega < f$ when $f_{eff} \downarrow 0$, which precludes observed vertical stratified-water motions at $\omega > f$.

Since the observed polarization switches are not related with environmental conditions like vertical N and $f_{eff}$, they may be related with turbulence intensity and source. While near-homogeneous conditions are characterized by general geothermal convection turbulence from below, the turbulence under stratified-water conditions varies between weakly turbulent interfacial instabilities and vigorous turbulent convection from above/sideways pushing stratification close to the seafloor (van Haren *et al*. 2026). Apparently, the first two turbulence processes occur at and limit energy transfer to frequencies not affecting inertial motions, while the vigorous internal-wave-induced convection is too fast for rotational near-inertial motions to develop resulting in a broad IGW with variable polarization and a negligible near-inertial peak in kinetic energy. The broad IGW smoothly extends into the sub-mesoscale band which also shows increased kinetic energy levels, like observed in near-equatorial as opposed to sub-tropical waters (van Haren 2005). These results indicate the importance of polarization variations over short distances of oscillatory motions for the redistribution of suspended mater and nutrients in different deep-sea environments.

**Acknowledgements.** Captains and crews of R/V Pelagia are thanked for the very pleasant cooperation. I also thank NIOZ colleagues notably from the NMF department for their indispensable contributions during the long preparatory and construction phases to make the unique sea-operation successful. I highly appreciated working with colleagues within the KM3NeT collaboration, including the stimulating discussions with E. Berbee on solving technical problems. The author acknowledges the partial financial support of Nederlandse organisatie voor Wetenschappelijk Onderzoek (NWO), the Netherlands.

**Declaration of interests.** The author declares no conflict of interest relevant to this study.

REFERENCES

Adrián-Martínez, S., Ageron, M., Aharonian, F., Aiello, S., Albert, A., *et al*. 2016 Letter of intent for KM3NeT 2.0. *J. Phys. G* **43**, 084001.

Bolgiano, R. 1959 Turbulent spectra in a stably stratified atmosphere. *J. Geophys. Res*. **64**, 2226-2229.

Briscoe, M.G. 1975 Preliminary results from the trimoored internal wave experiment (IWEX). *J. Geophys. Res*. **80**, 3872-3884.

Crepon, M., Wald, L. & Monget, J. M. 1982 Low-frequency waves in the Ligurian Sea during December 1977. *J. Geophys. Res*. **87**, 595-600.

Gargett, A.E., Hendricks, P.J., Sanford, T.B., Osborn, T.R. & Williams, A.J. 1981 A composite spectrum of vertical shear in the upper ocean. *J. Phys. Oceanogr*. **11**, 1258-1271.

Gascard, J.-C. 1978 Mediterranean deep water formation, baroclinic eddies and ocean eddies. *Oceanol. Acta* **1**, 315-330.

Gerkema, T. & Exarchou, E. 2008 Internal-wave properties in weakly stratified layers. *J. Mar. Res*. **66**, 617-644.

Gerkema, T., Zimmerman, J.T.F., Maas, L.R.M. & van Haren, H. 2008 Geophysical and astrophysical fluid dynamics beyond the traditional approximation. *Rev. Geophys*. **46**, RG2004.

Gonella, J. 1972 A rotary-component method for analyzing meteorological and oceanographic vector time series. *Deep Sea Res*. **19**, 833-846.

Kolmogorov, A.N. 1941 The local structure of turbulence in incompressible viscous fluid for very large Reynolds numbers. *Dokl. Akad. Nauk SSSR* **30**, 301-305.

Kunze, E. 1985 Near-inertial wave propagation in geostrophic shear. *J. Phys. Oceanogr*. **15**, 544-565.

LeBlond, P.H. & Mysak, L.A. 1978 Waves in the Ocean, Elsevier, New York, USA, 602 pp.

Lohse, D. & Xia, K.-Q. 2010 Small-Scale properties of turbulent Rayleigh-Bénard convection. *Annu. Rev. Fluid Mech*. **42**, 335-364.

Maas, L.R.M. & van Haren, J.J.M. 1987 Observations on the vertical structure of tidal and inertial currents in the central North Sea. *J. Mar. Res*. **45**, 293-318.

Marshall, J., & Schott, F. 1999 Open-ocean convection: Observations, theory, and models. *Rev. Geophys*. **37**, 1-64.

Mooers, C.N.K. 1975 Several effects of a baroclinic current on the cross-stream propagation of inertial-internal waves. *Geophys. Fluid Dyn*. **6**, 245-275.

Parks, T.W. & Burrus, C.S. 1987 Digital filter design, John Wiley & Sons, New York, USA, 342 pp.

Prandle, D. 1982 The vertical structure of tidal currents. *Geophys. Astrophys. Fluid Dyn*. **22**, 29-49.

Saunders, P.M. 1983 Benthic observations on the Madeira abyssal plain: Currents and dispersion. *J. Phys. Oceanogr*. **13**, 1416-1429.

Sheremet, V.A. 2004 Laboratory experiments with tilted convective plumes on a centrifuge: A finite angle between the buoyancy force and the axis of rotation. *J. Fluid Mech*. **506**, 217-244

Straneo, F., Kawase, M. & Riser, S.C. 2002 Idealized models of slantwise convection in a baroclinic flow. *J. Phys. Oceanogr*. **32**, 558-572.

Taylor, J.R. & Thompson, A.F. 2023 Submesoscale dynamics in the upper ocean. *Annu. Rev. Fluid Mech*. **55**, 103-127.

Testor, P. & Gascard, J.C. 2006 Post-convection spreading phase in the Northwestern Mediterranean Sea. *Deep-Sea Res*. **53**, 869-893.

van Haren, H. 2000 Properties of vertical current shear across stratification in the central North Sea. *J. Mar. Res*. **58**, 465-491.

van Haren, H. 2005 Sharp near-equatorial transitions in inertial motions and deep-ocean step-formation. *Geophys. Res. Lett*. **32**, L01605.

van Haren, H. 2015 Atypical anticlockwise internal tidal motions in the deep ocean. *Tellus A* **67**, 27718.

van Haren, H., 2025 Technical note: spectral slopes in deep, weakly-stratified ocean and coupling between sub-meoscale motions and small-scale mechanisms. *Ocean Sci.* **21**, 555-565.

van Haren, H. & Millot, C. 2004 Rectilinear and circular inertial motions in the Western Mediterranean Sea. *Deep-Sea Res. I* **51**, 1441-1455.

van Haren, H. & Millot, C. 2009 Slantwise convection: a candidate for homogenization of deep newly formed dense waters. *Geophys. Res. Lett*. **36**, L12604.

van Haren, H., Taupier-Letage I., Aguilar J.A., Albert A., Anghinolfi M., *et al*. 2011 Acoustic and optical variations during rapid downward motion episodes in the deep north-western Mediterranean Sea. *Deep-Sea Res. I* **58**, 875-884.

van Haren, H., Bakker, R., Witte, Y., Laan, M. & van Heerwaarden, J. 2021 Half a cubic hectometer mooring array 3D-T of 3000 temperature sensors in the deep sea. *J. Atmos. Ocean. Technol*. **38**, 1585-1597.

van Haren, H., Adriani, O., Albert, A., Alhebsi, A.R., Alshalloudi, S., *et al*. 2026 Whipped and mixed warm clouds in the deep sea. *Geophys. Res. Lett.* **53**, e2025GL119998.

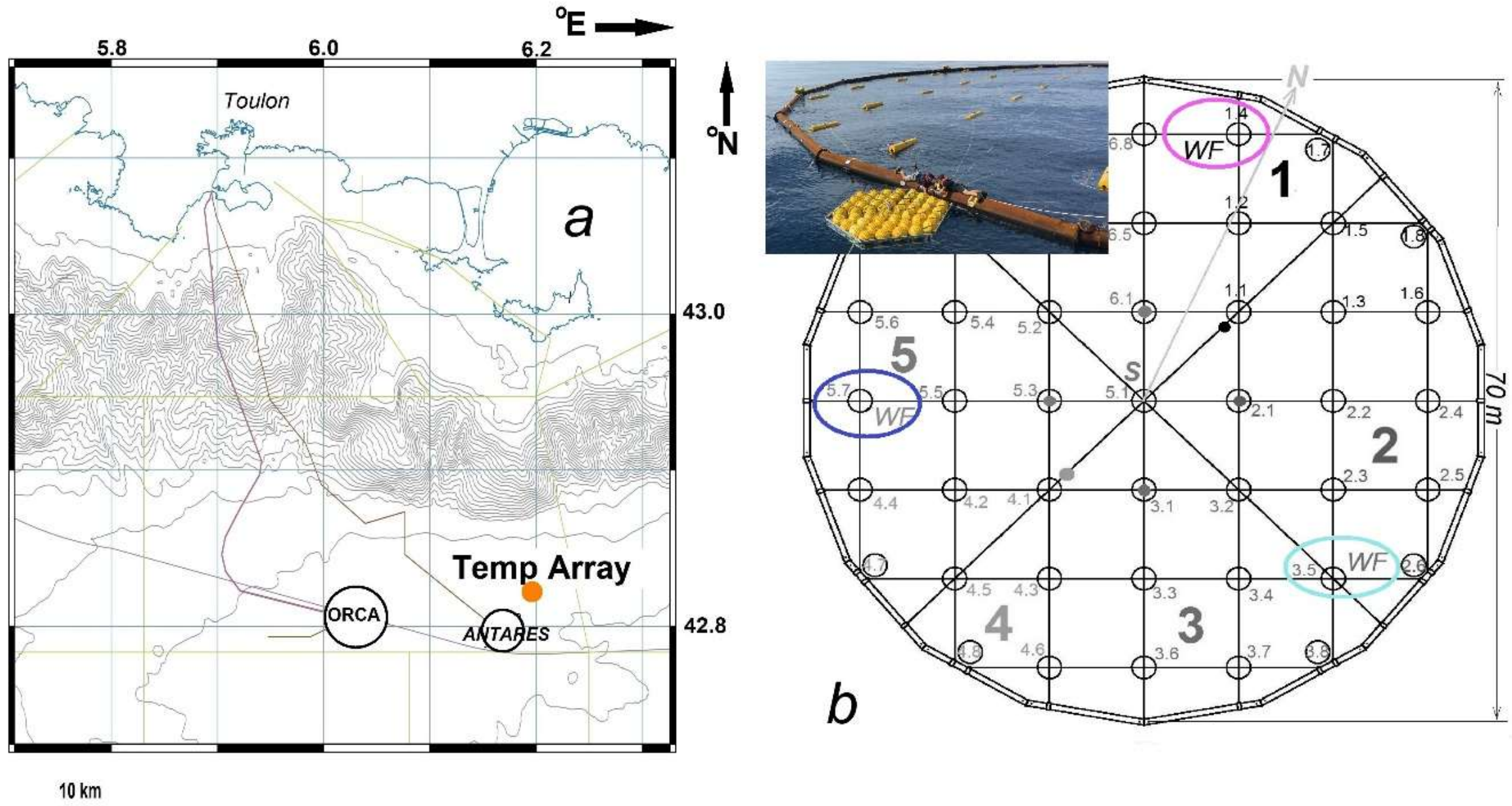

**Figure 1.** Mooring array location and layout. (a) Location named "Temp Array" (orange dot) on map off southern France. The mooring is well east of main neutrino-telescope site “ORCA” of KM3NeT collaboration (Adrián-Martinez *et al.* 2016) and just northeast of the former “ANTARES” neutrino-telescope site. Isobaths are drawn every 100 m. The grey lines indicate Toulon-harbour approach sectors. (b) Mooring-array layout and orientation at seabed, with steel-cable grid and small rings holding the vertical lines at 9.5 m horizontal intervals. Waterflow ‘WF’ current meters are at buoys on top of lines 14 (in magenta ellipse), 35 (cyan) and 57 (blue). The insert shows part of the large ring just prior to free-fall deployment at sea, with drag-parachute in front.

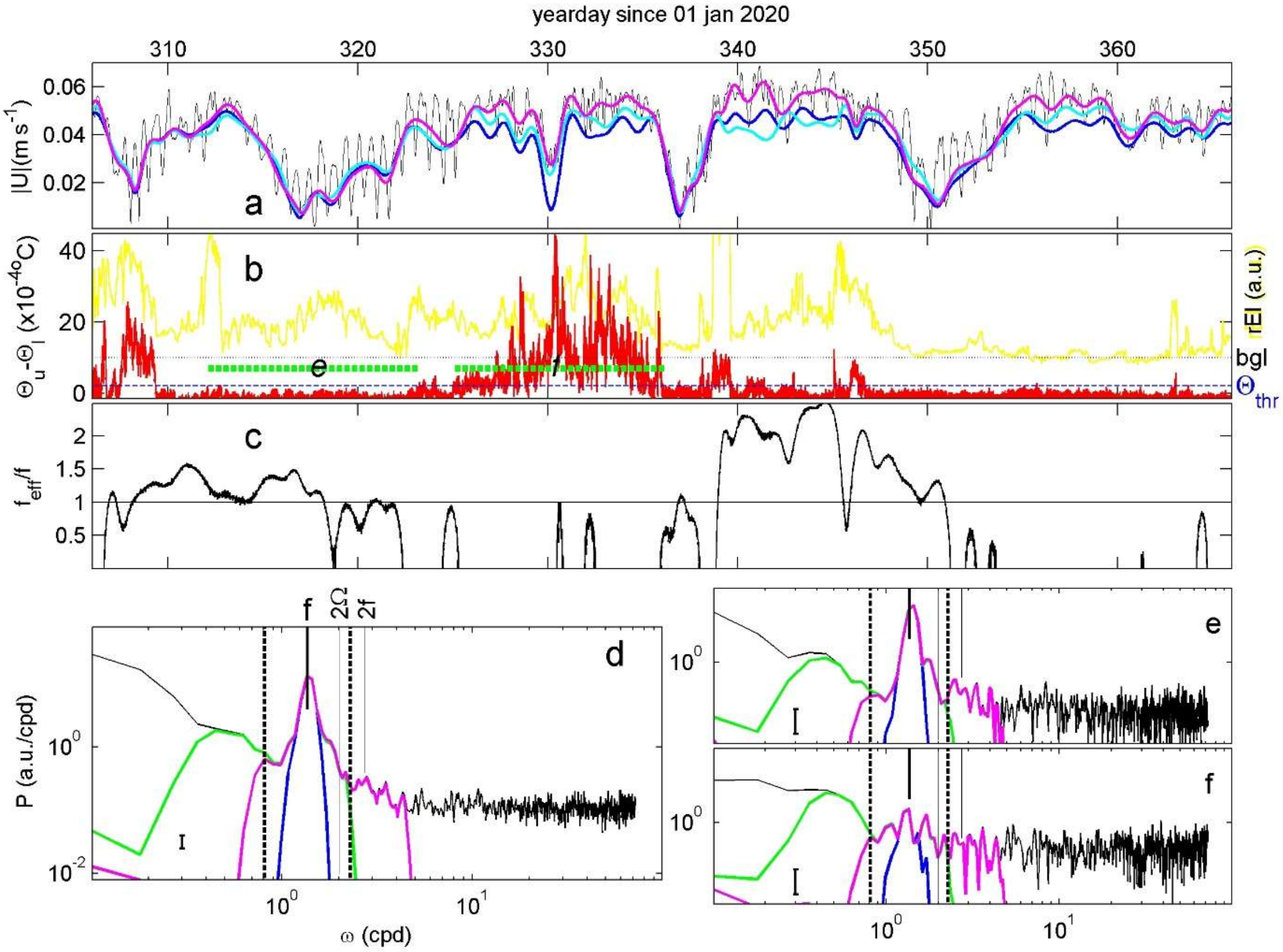

**Figure 2.** Two-month time series of current meter data. (a) Diurnally filtered (cut-off at 1 cpd, cycles per day) waterflow speed of three current meters (colour coding in Fig. 1b). In black, noise-filtered (cut-off at 4 cpd) data for line 14 (magenta). (b) Vertical temperature difference between uppermost and lowest T-sensor of line 17 (red) and relative echo intensity (yellow) averaged over the three current-meter records, with 'bgl' indicating the approximate background clear-water level. Horizontal dashed bars indicate periods of panels e. and f. (c) Sub-mesoscale and mesoscale effective inertial frequency relative to the local planetary inertial frequency. (d) Two-month averaged kinetic energy spectrum and filter cut-offs applied to line-57 data. The heavy vertical dashed lines indicate the inertio-gravity wave IGW bounds for buoyancy frequency N = 1.0f, 2Ω indicates the semidiurnal frequency. Original spectrum (black), IGW-$N_{max}$ (maximum small-scale) band with filter cut-offs at 0.8 and 4 cpd (magenta), sub-mesoscale IGW band with cut-offs at 0.4 and 2 cpd (green), and the near-inertial band with cut-offs at 0.87f and 1.13f (blue). (e) As d. but for 11-day period between days 312.0 and 323.2 under near-homogeneous conditions. (f) As e., but for days 325.0-336.2 under vertical stratified-water conditions.

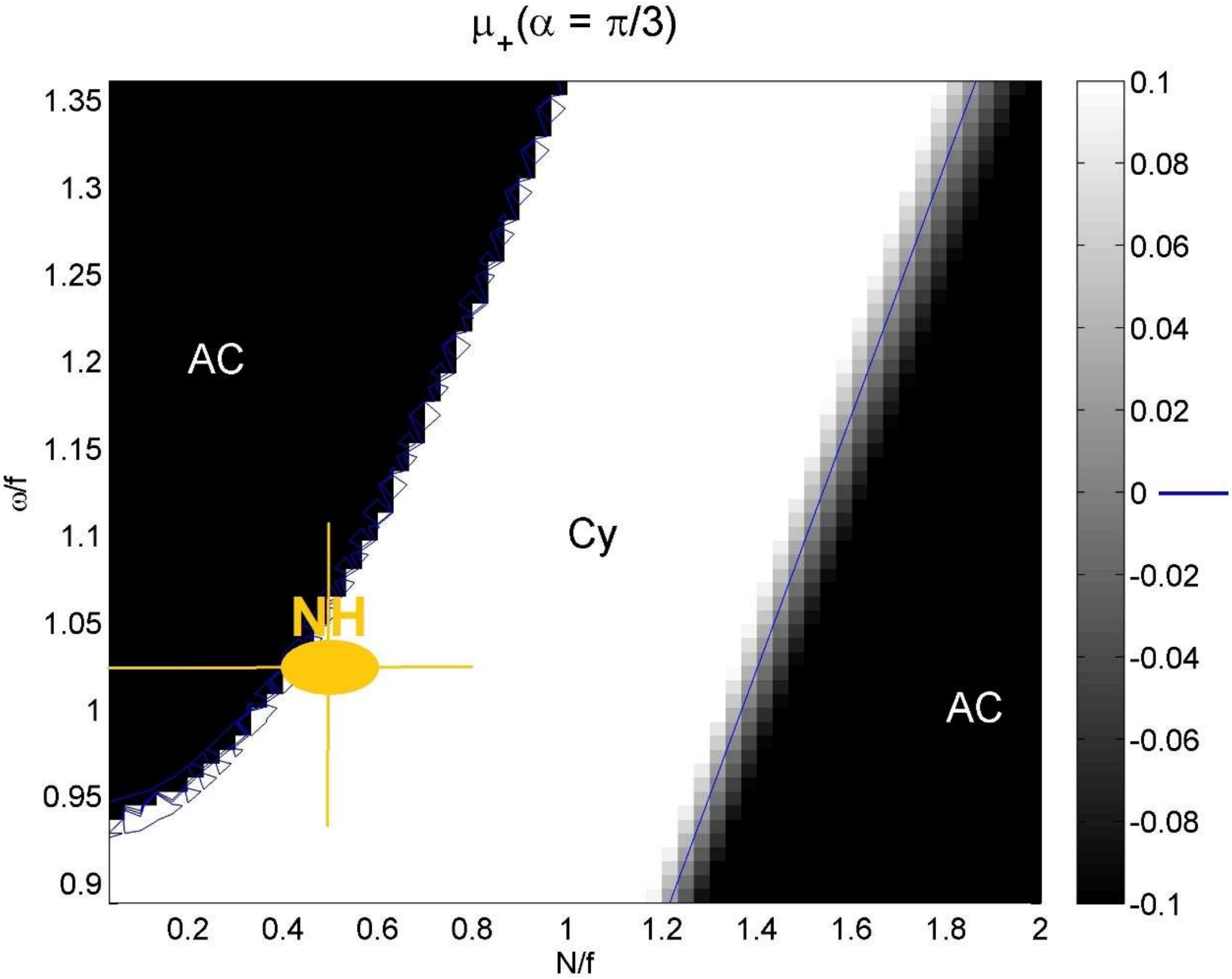

**Figure 3.** Parameter space (ω, N)/f of anticyclonic 'AC' and cyclonic 'Cy' polarization for horizontal near-inertial IGW motions under weak vertical density stratification with 0 < N < 2f that is typically observed in the deep Western Mediterranean. Based on model equations (2.3) and (2.4) for the upward directed beam in 60□-from-latitude direction. (For extensive description of theoretical model, the reader is referred to Gerkema & Excharchou 2008). Specifically, observed near-inertial peak frequencies and near-homogeneous 'NH' stratification ranges are indicated in yellow, with the cross reaching <10% occurrence values and the filled ellipse containing >60% of occurrence values.

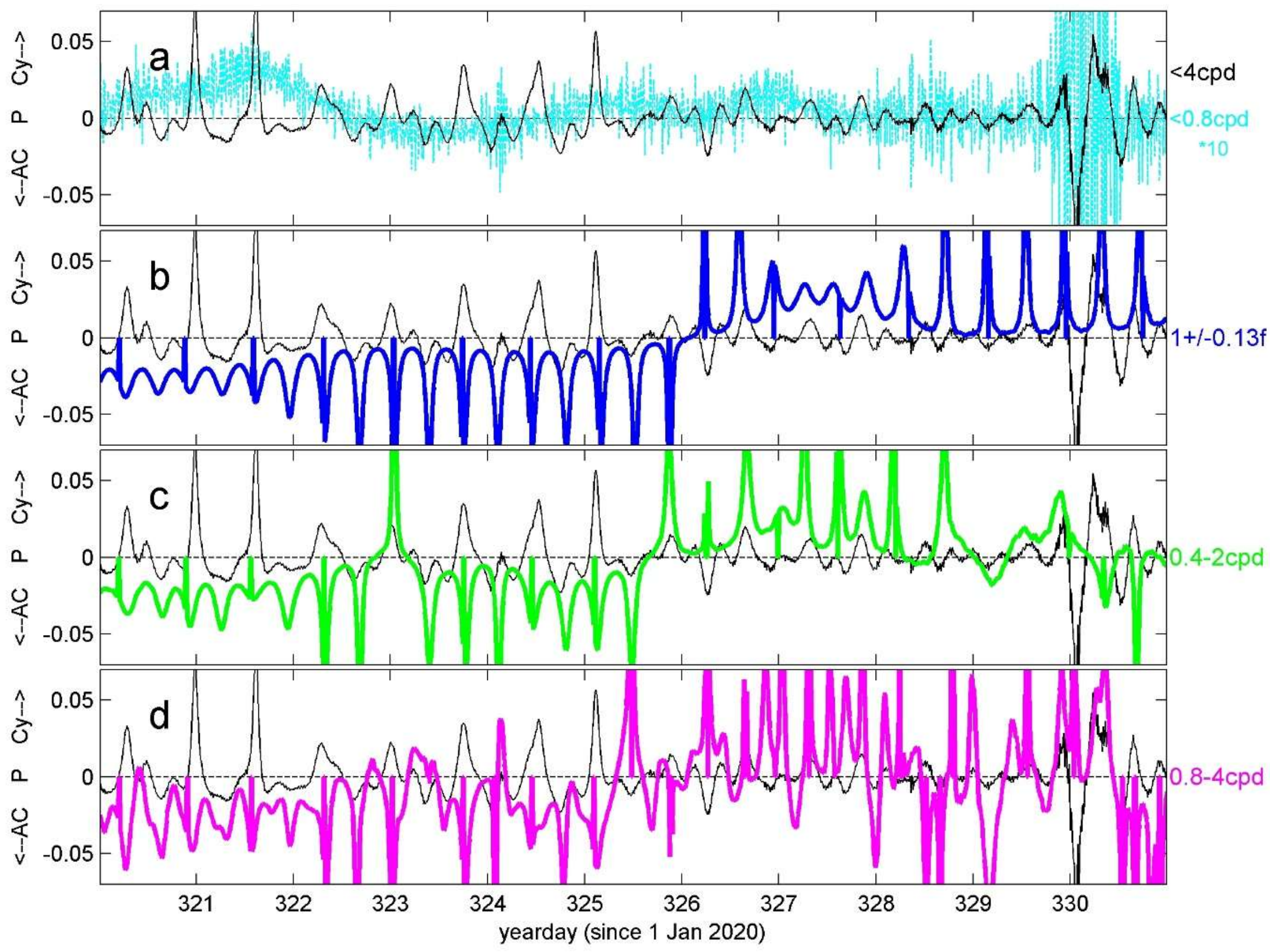

**Figure 4.** Eleven-day timeseries from current meter at line 57 of polarization (3.1), cyclonic 'Cy' when sign is positive, anticyclonic 'AC' when negative, during an episode of dominant eastward flow. (a) Noise filtered (background; black) and diurnally filtered data (x10; cyan). (b) As a., but for the near-inertial band. For filter bands see Fig. 2d. (c). As a., but for sub-mesoscale-IGW band. (d) As a., but for the IGW-$N_{max}$ band.

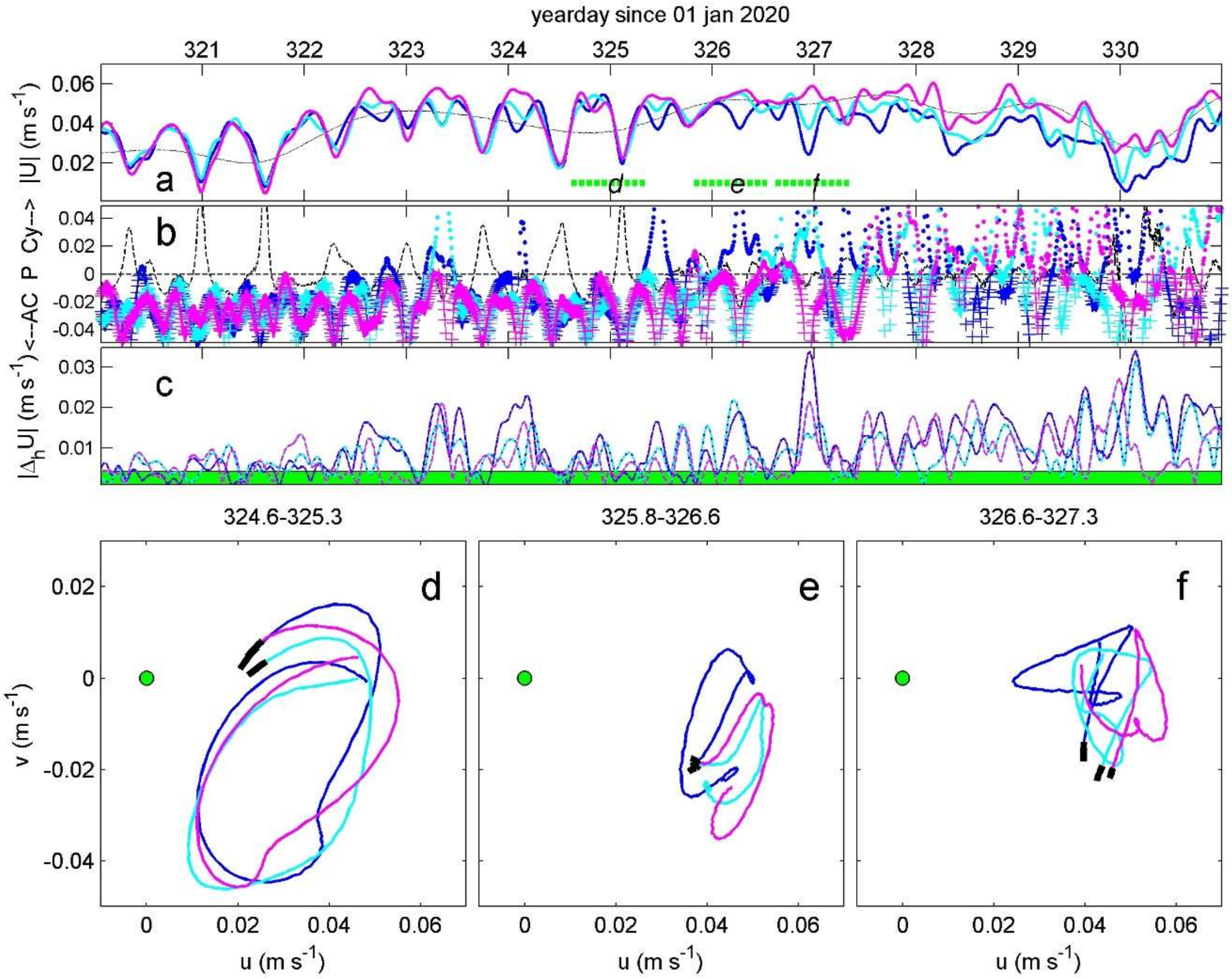

**Figure 5.** Polarization variations between the three current meters for the period of Fig. 4. (a). Noise-filtered data, with colour-coding of Fig. 1b. In thin black, diurnally-filtered data from line 14. The horizontal bars indicate the periods of panels d., e., and f. (b) Polarization sign for IGW-$N_{max}$ band data, with reference of line-57 noise-filtered data in black. (c) Amplitudes of horizontal flow differences between pairs of data in a. The height of the green bar indicates the 95% significance level. (d) One inertial period hodographs between indicated days. The start of each graph is in thick black. Episode with dominant cyclonic near-inertial motion for all three current meters. The green circle indicates one standard deviation error around (0, 0). (e) As d., but for an episode with mainly cyclonic line-57 data. (f) As d., but for an episode with mainly cyclonic line-35 and line-57 data, and short cyclonic line-14 data.